\documentclass[twocolumn,superscriptaddress,showpacs]{revtex4}

\begin{document}

\title{Implementation of qutrit-based quantum information processing via
state-dependent forces on trapped ions}
\author{Li-Xiang Cen}
\affiliation{Department of Physics, Sichuan University, Chengdu
610065, China}
\author{Bang-Pin Hou}
\affiliation{Department of Physics, Sichuan Normal University,
Chengdu 610068, China}
\author{Ming-Lun Chen}
\affiliation{Department of Physics, Sichuan University, Chengdu
610065, China}

\begin{abstract}
We propose a scheme to realize quantum logic and entanglement for
qutrit systems via state-dependent forces on trapped ions. By
exploiting the laser-ion coupling in the presence of Coulomb
interactions, the set of quantum gate operations including the
conditional phase shifts on two qutrits as well as arbitrary SU(3)
rotations on single qutrits are derived for universal quantum
manipulation. As an illustration, we demonstrate in detail how these
gate resources could be used to generate the maximally entangled
state of two qutrits. Besides being insensitive to vibrational
heating of the trapped ions, the present scheme is also shown to be
scalable through designing appropriately the pulse configuration of
the laser-ion interactions.
\end{abstract}


\maketitle

Quantum information processing based on trapped ions has witnessed
rapid development in the past decade. Since the first gate scheme
proposed by Cirac and Zoller \cite{cirac}, several scenarios have
been proposed for entangling trapped ion quantum bits for quantum
information applications
\cite{molmer,knight,push,wei,garcia,duan,zhu,cen}. It was shown
recently \cite{push,garcia,duan,zhu,cen} that quantum manipulation
could be implemented via state-dependent forces acting on ions in
the presence of mutual Coulomb interactions. Experimental studies
exploiting optical Raman fields to achieve strong state-dependent
forces through ac-Stark shifts have also been reported
\cite{leibfried,haljan}. Remarkably, owing to the global property of
the evolution in the position-momentum phase space, quantum
operations based on this kind of laser-ion coupling are insensitive
to vibrational heating from noisy background electric fields,
therefore relax the physical constraints on implementing quantum
information processing.

The conventional building blocks for quantum processors are qubits,
i.e., the two-level physical systems. On the other hand, the use of
quantum entanglement in high-dimensional quantum systems has also
been considered to generalize and improve the qubit-based quantum
algorithms and protocols \cite {carlton}. It was proven that
utilization of qutrit systems instead of qubits could be more secure
against a symmetric attack on a quantum key distribution protocol
\cite{bruss}. Furthermore, quantum computing power based on qudit
quantum processors has been investigated recently and the
qutrit-based quantum information processing is found to optimize the
Hilbert-space dimensionality hence maximize the computing power
\cite{green}. Motivated by these facts, designs of quantum
manipulation on qutrit systems including the direct extension of the
original Cirac-Zoller scenario have also been investigated
\cite{klimov,hugh}.

In this paper we shall present a scheme to realize the qutrit-based
quantum manipulation by making use of the state-dependent laser-ion
coupling. The derived set of quantum logical operations, including
the conditional phase shifts on two qutrits and arbitrary SU(3)
rotations on single qutrits, promises the universal processing of
quantum information. As an illustration, the maximally entangled
state of two qutrits is consequently generated by using these gate
resources. The built scheme of quantum manipulation is believed to
be insensitive to vibrational temperature of ions. Moreover, the
potential to scale up the ion array system is also demonstrated
through designing appropriately the pulse configuration of the
laser-ion interactions.

Let us consider a system of $N$ ions confined in a one-dimensional
linear Paul trap with a global trap frequency $\omega $. The
hyperfine internal states of each ion, \{$|0\rangle ,|1\rangle
,|2\rangle $\}, are selected to encode information for qutrits. In
our scheme, the rotations between any two levels of single qutrits,
namely, the transformations $R_{mm^{\prime }}(\theta ,\phi )$
specified by
\begin{eqnarray}
R_{mm^{\prime }}(\theta ,\phi )|m\rangle &=&\cos \theta |m\rangle +ie^{i\phi
}\sin \theta |m^{\prime }\rangle ,  \nonumber \\
R_{mm^{\prime }}(\theta ,\phi )|m^{\prime }\rangle &=&ie^{-i\phi }\sin
\theta |m\rangle +\cos \theta |m^{\prime }\rangle  \label{sgloprt}
\end{eqnarray}
with \{$|m\rangle ,|m^{\prime }\rangle $\} denoting any one of the
sets \{$|0\rangle ,|1\rangle $\}, \{$|0\rangle ,|2\rangle $\} and
\{$|1\rangle ,|2\rangle $\}, are performed without involving the
motional degree of freedom of ions. This is attainable in physics
either by applying resonant microwave fields to induce coherent Rabi
oscillations between the states $ |m\rangle $ and $|m^{\prime
}\rangle $, or by exploiting optical Raman fields to generate the
stimulated-Raman transitions through coupling virtually the excited
states \cite{leibried2}. Note that the rotations (\ref {sgloprt})
account for a series of SU(2) transformations that form the
necessity ingredients of the general SU(3) operation for single
qutrits.

The central issue we are going to discuss is the implementation of
conditional logic for two qutrits, in which the role of the vibrational mode
is necessary as it effectively couples the ions together in the presence of
Coulomb interactions. In the absence of external forces, the ion motion is
well characterized by the collective harmonic oscillators under the
second-order expansion of the potential for small vibrations. For
concreteness, we will focus on quantum operations on $N=2$ ions, and leave
the extension to large number of ions for a late discussion. The base
Hamiltonian for the system is therefore described as (setting $\hbar =1$)
\begin{equation}
H_0=\sum_{k=c,r}\omega _ka_k^{\dagger }a_k+\sum_{\mu =1,2}[\Delta _1\sigma
_{11}^{(\mu )}+\Delta _2\sigma _{22}^{(\mu )}],  \label{hamil0}
\end{equation}
where $\omega _c=\omega $ and $\omega _r$ denote the frequencies
associated with the center-of-mass and stretch modes, respectively;
and $a_k(a_k^{\dagger })$ are their corresponding annihilation
(creation) operators. The notation $\sigma _{mm^{\prime }}^{(\mu
)}=|m\rangle \langle m^{\prime }|$ ($m,m^{\prime }=0,1,2$) accounts
for internal level operators of the ion $\mu $, and $\Delta _1$ and
$\Delta _2$ are energy differences between the levels $|0\rangle
\leftrightarrow |1\rangle $ and $|1\rangle \leftrightarrow |2\rangle
$, respectively. To perform quantum gates, we assume that the ions
could be addressed individually by laser beams and the acceleration
force is hence created depending on the ion internal state, namely,
$F_\mu (t)=\sum_mf_{\mu ,m}(t)\sigma _{mm}^{(\mu )}$ with
coefficients $f_{\mu ,m}(t)$ specified by the detailed laser-ion
coupling \cite{wineland}. The related interaction term is given by
$H_F^{(\mu )}(t)=-x_\mu F_\mu (t)$, where the local coordinator
$x_\mu $ ($\mu =1,2$) relates to the collective one by
$q_c=(x_1+x_2)/\sqrt{2}$ and $q_r=(x_1-x_2)/\sqrt{2}$. As the two
ions are exerted by the external acceleration forces simultaneously,
the Hamiltonian of the system in the rotation frame with respect to
$H_0$ takes the following general form
\begin{equation}
H_I(t)=-\sum_{k,m}[g_{1,m}^k(t)\sigma _{mm}^{(1)}+g_{2,m}^k(t)\sigma
_{mm}^{(2)}]a_ke^{-i\omega _kt}+h.c..  \label{hamil}
\end{equation}
Here, the sums are taken over $k=c,r$ and $m=0,1,2$, and the coefficients
are given by $g_{\mu ,m}^k(t)=D_{\mu k}f_{\mu ,m}(t)/\sqrt{2M\omega _k}$ in
which $D$ is an orthogonal matrix $D=\frac 1{\sqrt{2}}$(\negthinspace {\tiny
$
\begin{array}{ll}
1 & 1 \\
1 & -1
\end{array}
\!\!$}) and $M$ denotes the mass of the ion.

To explore the time evolution of the system, we resort to a gauged
representation \cite{wangsj} with respect to the unitary
transformation $G(t)=\exp [-i\int_0^tH_I(\tau )d\tau ]$. It is
directly shown that the time evolution operator of the system reads
$U(t)=G(t)U_g(t)$, where $U_g(t)$ satisfies the covariant equation
$i\partial _tU_g(t)=H_I^g(t)U_g(t)$ and the gauged Hamiltonian
$H_I^g(t)$ is worked out to be
\begin{eqnarray}
H_I^g(t) &=&G^{-1}H_IG-iG^{-1}\partial G/\partial t  \nonumber \\
&=&\sum_{m,n}J_{mn}(t)\sigma _{mm}^{(1)}\sigma _{nn}^{(2)}  \nonumber \\
&&+\sum_m[\epsilon _{1,m}(t)\sigma _{mm}^{(1)}+\epsilon _{2,m}(t)\sigma
_{mm}^{(2)}],  \label{gaugeh}
\end{eqnarray}
where the coefficients $J_{mn}(t)$ and $\epsilon _{\mu ,m}(t)$ ($\mu =1,2$)
are given by
\begin{eqnarray}
J_{mn}(t) &=&\sum_{k=c,r}\int_0^t[g_{1,m}^k(t)g_{2,n}^k(t^{\prime
})+g_{1,m}^k(t^{\prime })g_{2,n}^k(t)]  \nonumber \\
&&\times \sin \omega _k(t^{\prime }-t)dt^{\prime },  \nonumber \\
\epsilon _{\mu ,m}(t) &=&\sum_{k=c,r}\int_0^tg_{\mu ,m}^k(t)g_{\mu
,m}^k(t^{\prime })\sin \omega _k(t^{\prime }-t)dt^{\prime }.  \label{coupj}
\end{eqnarray}
It is seen that the pure qutrit-qutrit coupling is explicitly
manifested in the gauged Hamiltonian of Eq. (\ref{gaugeh}), and the
coupling of the qutrit with phonon degree of freedom is only
indicated by the transformation $G(t)$. Especially, as the external
force is controlled with a particular configuration such that
\begin{equation}
\int_0^Tg_{\mu ,m}^k(t)e^{-i\omega _kt}dt=0,~~k=c,r,  \label{config}
\end{equation}
the transformation $G(T)$ becomes an identity operator. Consequently, the
evolution operator of the system at time $T$ is exactly obtained as
\begin{eqnarray}
U(T) &=&U_g(T)  \nonumber \\
&=&\left[ \prod_{m,n=0}^2P_{mn}\right] \left[ \prod_{m=0}^2D_m^{(1)}\right]
\left[ \prod_{m=0}^2D_m^{(2)}\right] ,  \label{evolution}
\end{eqnarray}
where
\begin{equation}
P_{mn}\equiv e^{-i\Phi _{mn}(T)\sigma _{mm}^{(1)}\sigma
_{nn}^{(2)}},~~D_m^{(\mu )}\equiv e^{-i\phi _{\mu ,m}(T)\sigma _{mm}^{(\mu
)}},  \label{defin}
\end{equation}
and the coefficients are given by
\begin{equation}
\Phi _{mn}(T)=\int_0^TJ_{mn}(t)dt,\phi _{\mu ,m}(T)=\int_0^T\epsilon _{\mu
,m}(t)dt.  \label{parame}
\end{equation}

The intriguing feature of the above process is that the generated
overall evolution (\ref{evolution}) contains no operator entangling
the qutrit with vibrational degrees of freedom, hence is in sense
global and fault tolerant against the vibrational heating of the
ions. The first factor contained in the expression of
(\ref{evolution}) stands explicitly for a general phase shift
operation for two qutrits. The derived form is actually universal in
view that it exhausts all of the possible two-qutrit logic
consisting of phase flip $P_{mn}(\Phi )$, providing that appropriate
selections of the laser-ion coupling are promised. It is seen that
the induced transformation (\ref{evolution}) includes also two extra
factors composed of operators $D_m^{(\mu )}$ acting on the two
qutrits individually. Differing from the case of the qubit system,
these extra operations are inevitable in the evolution specified by
Eqs. (\ref{gaugeh}) and (\ref{coupj}) and indicate excessive actions
in the above process to achieve the desired two-qutrit phase shift
operation. To remove these undesirable actions, we present in the
following a scheme to achieve the independent phase shift operation
for single qutrits which combining with the transformation
(\ref{evolution}) could induce the pure logical operations for two
qutrits.

In detail, we exploit the same physical setup to exert the acceleration
force on the individual ion $\mu $, by which the Hamiltonian in the rotation
frame reads
\begin{equation}
H_I^{(\mu )}(t)=-\sum_{k=c,r}[g_m^k(t)a_ke^{-i\omega _kt}+h.c.]\sigma
_{mm}^{(\mu )},  \label{singh}
\end{equation}
where $m=0,1,$ or $2$. As the system undergoes a cyclic evolution
such that $\int_0^Tg_m^k(t)e^{-i\omega _kt}dt=0$, the evolution
operator of the system is obtained as
\begin{equation}
U(T)=e^{-i\phi (T)\sigma _{mm}^{(\mu )}}\equiv D_m^{(\mu )}(\phi ),
\label{singop}
\end{equation}
where
\begin{equation}
\phi (T)=\sum_{k=c,r}\int_0^T\int_0^tg_m^k(t)g_m^k(t^{\prime })\sin \omega
_k(t^{\prime }-t)dt^{\prime }dt.  \label{phase}
\end{equation}
Physically, the ion $\mu $ is displaced coherently in the evolution
if the electron inhabits in the level $|m\rangle $. By designing the
cyclic force configuration such that the ion returns to its original
state in the phase-momentum space, the vibrational degree of freedom
will become disentangled with the qutrit and an effective phase
shift independent of the motional state hence is induced for the ion
internal state. Note that there is the relation $\sum_m\sigma
_{mm}^{(\mu )}\equiv I$, hence for each ion only two of the
operators in (\ref{singop}) are independent. The significance of the
phase shift operations (\ref{singop}) derived for single qutrits is
actually multi-fold. Firstly, by noticing that all the operators
$D_m^{(1)}$, $D_m^{(2)}$, and $P_{mn}$ are commutative with each
other, the independent manipulation of $D_m^{(\mu )}(\phi )$
therefore provides a distinct way to remove the extra operations in
(\ref{evolution}) so that the pure conditional phase gate for two
qutrits could be achieved. Moreover, it is noteworthy that the two
independent phase gates of (\ref{singop}) enable the scheme,
together with the series of SU(2) rotations indicated in Eq.
(\ref{sgloprt}), to perform universal rotations for single qutrits.
Finally, we point out that the state evolution described above is
expressed in the rotation frame. It differs from the original
Schrodinger picture by a transformation $e^{-iH_0T}$ which would
induce relative phase accumulations to the qutrit states. With the
help of the phase-shift operation (\ref{singop}), the accumulation
of additional phases could be in principle canceled out.

Up to now, we have shown that the universal rotation for single
qutrits as well as the two-qutrit phase shift operation could be
achieved by virtue of the state-dependent forces on trapped ions.
These gate resources are sufficient to generate universal quantum
manipulations for qutrit systems. In particular, we show in the
following the utilization of the proposed gate operations to achieve
the maximally entangled state for two qutrits. In detail, assume
that the two qutrits are initially prepared in a state $|0\rangle
\otimes |0\rangle $. We first exert the same rotation on each qutrit
individually to derive a state $|\psi \rangle =|+\rangle \otimes
|+\rangle $, where
\begin{eqnarray}
|+\rangle  &=&R_{12}(\pi /4,-\pi /2)R_{01}(\arctan \sqrt{2},-\pi
/2)|0\rangle   \nonumber \\
&=&\frac 1{\sqrt{3}}(|0\rangle +|1\rangle +|2\rangle ).  \label{enop1}
\end{eqnarray}
Then by applying the sequence of two-qutrit phase flip operations we obtain
\begin{eqnarray}
|\psi _M^{\prime }\rangle  &=&P_{33}(2\pi /3)P_{22}(2\pi /3)P_{11}(2\pi
/3)|\psi \rangle   \nonumber \\
&=&\frac 1{\sqrt{3}}(|00^{\prime }\rangle +|11^{\prime }\rangle +|22^{\prime
}\rangle ),  \label{maxi}
\end{eqnarray}
where
\begin{eqnarray}
|0^{\prime }\rangle  &=&\frac 1{\sqrt{3}}(e^{-i2\pi /3}|0\rangle +|1\rangle
+|2\rangle ),  \nonumber \\
|1^{\prime }\rangle  &=&\frac 1{\sqrt{3}}(|0\rangle +e^{-i2\pi /3}|1\rangle
+|2\rangle ),  \nonumber \\
|2^{\prime }\rangle  &=&\frac 1{\sqrt{3}}(|0\rangle +|1\rangle +e^{-i2\pi
/3}|2\rangle ).  \label{locals}
\end{eqnarray}
In Eq. (\ref{maxi}) we have employed the aforementioned design using
joint evolution indicated by Eqs. (\ref{evolution}) and
(\ref{singop}) to achieve the pure phase shift operations
$P_{mm}(2\pi /3)$ with $m=1,2,3$. Consequently, the maximally
entangled state in the conventional computational basis, i.e.,
$|\psi _M\rangle =\frac 1{\sqrt{3}}\sum_{m=0}^2|mm\rangle $, could
be obtained by (up to an irrelevant global phase)
\begin{eqnarray}
&&R_{01}(\pi /4,5\pi /6)R_{02}(\arctan 1/\sqrt{2},5\pi /6)  \nonumber \\
&&\times R_{12}(\pi /4,\pi )D_1(\pi /6)|\psi _M^{\prime }\rangle \rightarrow
|\psi _M\rangle ,  \label{maxsta}
\end{eqnarray}
where the sequence of operations $R_{01}R_{02}R_{12}D_1$ are
performed on arbitrary one of the two qutrits.

The scheme we discussed till now is restricted to the system of two ions.
The extended version of the system with large $N$ ions should be described
by a Hamiltonian of form (\ref{hamil}) where in stead the sum of phonon
modes takes over $k=1,\cdots ,N$ and the coefficients are given by $g_{\mu
,m}^k=D_{\mu k}f_{\mu ,m}/\sqrt{2M\omega _k}$ with $D$ the unitary
transformation diagonalising the Hessian matrix. Note that the
commensurability condition (\ref{config}) is now replaced by the following
generalized one
\begin{equation}
\int_0^Tg_{\mu ,m}^k(t)e^{-i\omega _kt}dt=0,k=1,\cdots ,N  \label{gcond}
\end{equation}
which becomes very fragile due to the increased complexity of phonon
mode spectrum. One possible resolution to this difficulty is to
carry out the acceleration forces by an adiabatic manner.
Specifically, suppose that $f_{\mu ,m}(t)$ describing the
configuration of the pushing force is some smooth function and
satisfies $|\dot{f}_{\mu ,m}(t)|/\sqrt{2M\omega _c}\ll \omega _c$
where $\omega _c=\omega $ is the frequency of the center-of-mass
mode, i.e., the lowest frequency amongst all of the phonon modes
$\omega _k$. If the coefficient $g_{\mu ,m}^k(t)$ which is
proportional to $f_{\mu ,m}(t)$ undergoes from $0$ to some finite
value and then back to $0$ as time goes from $0$ to $T$, then all
the relations of (\ref{gcond}) come into existence
\begin{equation}
\int_0^Tg_{\mu ,m}^k(t)e^{-i\omega _kt}dt=\int_0^Ti[\dot{g}_{\mu
,m}^k(t)/\omega _k]e^{-i\omega _kt}dt\rightarrow 0.  \label{adiab}
\end{equation}
Moreover, the phases in Eq. (\ref{parame}) for this special case are
worked out to be
\begin{eqnarray}
\Phi _{mn}(T) &=&-2\sum_{k=1}^N\int_0^Tg_{1,m}^k(t)g_{2,n}^k(t)/\omega _kdt,
\nonumber \\
\phi _{\mu ,m}(T) &=&-\sum_{k=1}^N\int_0^T[g_{\mu ,m}^k(t)]^2/\omega _kdt.
\label{adiap}
\end{eqnarray}

The intrinsic drawback associated with the above adiabatic scenario
is the slow evolution of the gate operation which gives decoherence
more time to exert its detrimental effects. An alternative scaling
scenario is to employ the fast gate scheme combining with a noise
cancelation strategy of refocusing techniques \cite{duan,cen2}.
Briefly, the key point of the strategy is to observe that the
opposite loop evolution generated by reversal force configuration,
say, $f_{\mu ,m}^c(t)$ and $f_{\mu ,m}^{\bar{c}}(t)=-f_{\mu
,m}^c(t)$, would lead to the same $U_g(T)$ but reversed ingredients
of the noise contribution: $G_{\bar{c}}(T)=G_c^{-1}(T)$. Therefore
the influence of non-vanishing $G(T)$ could be effectively
suppressed by refocusing the gate pulses providing that the periodic
laser pulse is faster than all the frequencies of the vibrational
modes. For the further demonstration of validity of the scheme
including numerical illustration of the gate infidelity in the
presence of thermal phonon excitations, as the detailed discussion
for ion qubits was presented in Ref. \cite{duan}, we point out that
the same argument is valid for the present scenario of extended
qutrit systems.

In summary, we have proposed a scheme to implement the qutrit-based
quantum information processing by making use of the state-dependent
laser-ion coupling. The derived quantum logical operations are
believed to possess the noise resilience property against
vibrational heating of ions owing to the global feature of the
evolution in the phase space. We present also a detailed
illustration to generate the maximally entangled state of two
qutrits by utilizing these gate resources. Finally, the potential to
scale up the gate scheme for large-scale information processing is
also demonstrated, either through designing an adiabatic way of
switching on and off the interactions, or by a fast gate scenario
with refocusing strategy to cancel out the unwanted noise influence.

Support from the Youth Science Research Grants of Sichuan University is
gratefully acknowledged.

\end{document}